\documentclass[aps,
  prl,
  twocolumn,
  showpacs,
  superscriptaddress
]{revtex4-1}
              
\usepackage{color}
\usepackage{graphicx}
\usepackage{latexsym}
\usepackage{float}
\usepackage{amsmath}
\usepackage{amssymb}
\usepackage{wasysym}
\usepackage{multirow}
\usepackage{epsfig}
\usepackage[utf8]{inputenc}

\def\unnumberedsection{\unnumberedsectionwithoneargument}
\def\unnumberedsectionwithoneargument#1{
  \vskip 2ex\goodbreak
  \noindent
  \leavevmode
  \begingroup
  \bfseries
  #1.\quad
  \endgroup
}
    
\begin{document}


\title{Constraints on Cosmic Strings from the LIGO-Virgo Gravitational-Wave Detectors.}

\begin{abstract}
  Cosmic strings can give rise to a large variety of interesting
  astrophysical phenomena. Among them, powerful bursts of
  gravitational waves (GWs) produced by cusps are a promising
  observational signature. In this Letter we present a search for GWs
  from cosmic string cusps in data collected by the LIGO and Virgo
  gravitational wave detectors between 2005 and 2010, with over 625
  days of live time. We find no evidence of GW signals from cosmic
  strings. From this result, we derive new constraints on cosmic
  string parameters, which complement and improve existing limits from
  previous searches for a stochastic background of GWs from cosmic
  microwave background measurements and pulsar timing data. In
  particular, if the size of loops is given by the gravitational
  backreaction scale, we place upper limits on the string tension
  $G\mu$ below $10^{-8}$ in some regions of the cosmic string
  parameter space.
\end{abstract}

\author{%
J.~Aasi$^{1}$,
J.~Abadie$^{1}$,
B.~P.~Abbott$^{1}$,
R.~Abbott$^{1}$,
T.~Abbott$^{2}$,
M.~R.~Abernathy$^{1}$,
T.~Accadia$^{3}$,
F.~Acernese$^{4,5}$,
C.~Adams$^{6}$,
T.~Adams$^{7}$
% ...
\\
{{}$^{\dag}$Deceased, April 2012.}
{{}$^{\ddag}$Deceased, May 2012.}
}\noaffiliation

\affiliation {LIGO - California Institute of Technology, Pasadena, CA 91125, USA }
\affiliation {Louisiana State University, Baton Rouge, LA 70803, USA }
\affiliation {Laboratoire d'Annecy-le-Vieux de Physique des Particules (LAPP), Universit\'e de Savoie, CNRS/IN2P3, F-74941 Annecy-le-Vieux, France }
\affiliation {INFN, Sezione di Napoli, Complesso Universitario di Monte S.Angelo, I-80126 Napoli, Italy }
\affiliation {Universit\`a di Salerno, Fisciano, I-84084 Salerno, Italy }
\affiliation {LIGO - Livingston Observatory, Livingston, LA 70754, USA }
\affiliation {Cardiff University, Cardiff, CF24 3AA, United Kingdom }
\affiliation {Albert-Einstein-Institut, Max-Planck-Institut f\"ur Gravitationsphysik, D-30167 Hannover, Germany }
\affiliation {Nikhef, Science Park, 1098 XG Amsterdam, The Netherlands }
\affiliation {LIGO - Massachusetts Institute of Technology, Cambridge, MA 02139, USA }
\affiliation {Instituto Nacional de Pesquisas Espaciais, 12227-010 - S\~{a}o Jos\'{e} dos Campos, SP, Brazil }
\affiliation {University of Wisconsin--Milwaukee, Milwaukee, WI 53201, USA }
\affiliation {Leibniz Universit\"at Hannover, D-30167 Hannover, Germany }
\affiliation {INFN, Sezione di Pisa, I-56127 Pisa, Italy }
\affiliation {Universit\`a di Siena, I-53100 Siena, Italy }
\affiliation {University of Florida, Gainesville, FL 32611, USA }
\affiliation {The University of Mississippi, University, MS 38677, USA }
\affiliation {California State University Fullerton, Fullerton, CA 92831, USA }
\affiliation {INFN, Sezione di Roma, I-00185 Roma, Italy }
\affiliation {University of Birmingham, Birmingham, B15 2TT, United Kingdom }
\affiliation {Albert-Einstein-Institut, Max-Planck-Institut f\"ur Gravitationsphysik, D-14476 Golm, Germany }
\affiliation {Montana State University, Bozeman, MT 59717, USA }
\affiliation {European Gravitational Observatory (EGO), I-56021 Cascina, Pisa, Italy }
\affiliation {Syracuse University, Syracuse, NY 13244, USA }
\affiliation {LIGO - Hanford Observatory, Richland, WA 99352, USA }
\affiliation {SUPA, University of Glasgow, Glasgow, G12 8QQ, United Kingdom }
\affiliation {APC, AstroParticule et Cosmologie, Universit\'e Paris Diderot, CNRS/IN2P3, CEA/Irfu, Observatoire de Paris, Sorbonne Paris Cit\'e, 10, rue Alice Domon et L\'eonie Duquet, F-75205 Paris Cedex 13, France }
\affiliation {Columbia University, New York, NY 10027, USA }
\affiliation {Stanford University, Stanford, CA 94305, USA }
\affiliation {Universit\`a di Pisa, I-56127 Pisa, Italy }
\affiliation {CAMK-PAN, 00-716 Warsaw, Poland }
\affiliation {INFN, Sezione di Genova, I-16146 Genova, Italy }
\affiliation {Universit\`a degli Studi di Genova, I-16146 Genova, Italy }
\affiliation {The University of Texas at Brownsville, Brownsville, TX 78520, USA }
\affiliation {San Jose State University, San Jose, CA 95192, USA }
\affiliation {Moscow State University, Moscow, 119992, Russia }
\affiliation {LAL, Universit\'e Paris-Sud, IN2P3/CNRS, F-91898 Orsay, France }
\affiliation {NASA/Goddard Space Flight Center, Greenbelt, MD 20771, USA }
\affiliation {University of Western Australia, Crawley, WA 6009, Australia }
\affiliation {Universit\'e Nice-Sophia-Antipolis, CNRS, Observatoire de la C\^ote d'Azur, F-06304 Nice, France }
\affiliation {Institut de Physique de Rennes, CNRS, Universit\'e de Rennes 1, F-35042 Rennes, France }
\affiliation {Laboratoire des Mat\'eriaux Avanc\'es (LMA), IN2P3/CNRS, Universit\'e de Lyon, F-69622 Villeurbanne, Lyon, France }
\affiliation {Washington State University, Pullman, WA 99164, USA }
\affiliation {INFN, Sezione di Perugia, I-06123 Perugia, Italy }
\affiliation {INFN, Sezione di Firenze, I-50019 Sesto Fiorentino, Firenze, Italy }
\affiliation {Universit\`a degli Studi di Urbino 'Carlo Bo', I-61029 Urbino, Italy }
\affiliation {University of Oregon, Eugene, OR 97403, USA }
\affiliation {Laboratoire Kastler Brossel, ENS, CNRS, UPMC, Universit\'e Pierre et Marie Curie, F-75005 Paris, France }
\affiliation {Astronomical Observatory Warsaw University, 00-478 Warsaw, Poland }
\affiliation {VU University Amsterdam, 1081 HV Amsterdam, The Netherlands }
\affiliation {University of Maryland, College Park, MD 20742, USA }
\affiliation {University of Massachusetts - Amherst, Amherst, MA 01003, USA }
\affiliation {Universitat de les Illes Balears, E-07122 Palma de Mallorca, Spain }
\affiliation {Universit\`a di Napoli 'Federico II', Complesso Universitario di Monte S.Angelo, I-80126 Napoli, Italy }
\affiliation {Canadian Institute for Theoretical Astrophysics, University of Toronto, Toronto, Ontario, M5S 3H8, Canada }
\affiliation {Tsinghua University, Beijing 100084, China }
\affiliation {University of Michigan, Ann Arbor, MI 48109, USA }
\affiliation {Rochester Institute of Technology, Rochester, NY 14623, USA }
\affiliation {INFN, Sezione di Roma Tor Vergata, I-00133 Roma, Italy }
\affiliation {National Tsing Hua University, Hsinchu Taiwan 300 }
\affiliation {Charles Sturt University, Wagga Wagga, NSW 2678, Australia }
\affiliation {Caltech-CaRT, Pasadena, CA 91125, USA }
\affiliation {Pusan National University, Busan 609-735, Korea }
\affiliation {Australian National University, Canberra, ACT 0200, Australia }
\affiliation {Carleton College, Northfield, MN 55057, USA }
\affiliation {INFN, Gran Sasso Science Institute, I-67100 L'Aquila, Italy }
\affiliation {Universit\`a di Roma Tor Vergata, I-00133 Roma, Italy }
\affiliation {Universit\`a di Roma 'La Sapienza', I-00185 Roma, Italy }
\affiliation {University of Sannio at Benevento, I-82100 Benevento, Italy and INFN (Sezione di Napoli), Italy }
\affiliation {The George Washington University, Washington, DC 20052, USA }
\affiliation {University of Cambridge, Cambridge, CB2 1TN, United Kingdom }
\affiliation {University of Minnesota, Minneapolis, MN 55455, USA }
\affiliation {The University of Sheffield, Sheffield S10 2TN, United Kingdom }
\affiliation {Wigner RCP, RMKI, H-1121 Budapest, Konkoly Thege Mikl\'os \'ut 29-33, Hungary }
\affiliation {Inter-University Centre for Astronomy and Astrophysics, Pune - 411007, India }
\affiliation {INFN, Gruppo Collegato di Trento, I-38050 Povo, Trento, Italy }
\affiliation {Universit\`a di Trento, I-38050 Povo, Trento, Italy }
\affiliation {California Institute of Technology, Pasadena, CA 91125, USA }
\affiliation {Northwestern University, Evanston, IL 60208, USA }
\affiliation {Montclair State University, Montclair, NJ 07043, USA }
\affiliation {The Pennsylvania State University, University Park, PA 16802, USA }
\affiliation {MTA-Eotvos University, \lq Lendulet\rq A. R. G., Budapest 1117, Hungary }
\affiliation {National Astronomical Observatory of Japan, Tokyo 181-8588, Japan }
\affiliation {Universit\`a di Perugia, I-06123 Perugia, Italy }
\affiliation {Rutherford Appleton Laboratory, HSIC, Chilton, Didcot, Oxon, OX11 0QX, United Kingdom }
\affiliation {Embry-Riddle Aeronautical University, Prescott, AZ 86301, USA }
\affiliation {Department of Astrophysics/IMAPP, Radboud University Nijmegen, P.O. Box 9010, 6500 GL Nijmegen, The Netherlands }
\affiliation {Perimeter Institute for Theoretical Physics, Ontario, N2L 2Y5, Canada }
\affiliation {American University, Washington, DC 20016, USA }
\affiliation {University of New Hampshire, Durham, NH 03824, USA }
\affiliation {College of William and Mary, Williamsburg, VA 23187, USA }
\affiliation {University of Adelaide, Adelaide, SA 5005, Australia }
\affiliation {Raman Research Institute, Bangalore, Karnataka 560080, India }
\affiliation {Korea Institute of Science and Technology Information, Daejeon 305-806, Korea }
\affiliation {Bia{\l }ystok University, 15-424 Bia{\l }ystok, Poland }
\affiliation {University of Southampton, Southampton, SO17 1BJ, United Kingdom }
\affiliation {IISER-TVM, CET Campus, Trivandrum Kerala 695016, India }
\affiliation {Hobart and William Smith Colleges, Geneva, NY 14456, USA }
\affiliation {Institute of Applied Physics, Nizhny Novgorod, 603950, Russia }
\affiliation {Seoul National University, Seoul 151-742, Korea }
\affiliation {Hanyang University, Seoul 133-791, Korea }
\affiliation {IM-PAN, 00-956 Warsaw, Poland }
\affiliation {NCBJ, 05-400 \'Swierk-Otwock, Poland }
\affiliation {Institute for Plasma Research, Bhat, Gandhinagar 382428, India }
\affiliation {Utah State University, Logan, UT 84322, USA }
\affiliation {The University of Melbourne, Parkville, VIC 3010, Australia }
\affiliation {University of Brussels, Brussels 1050 Belgium }
\affiliation {SUPA, University of Strathclyde, Glasgow, G1 1XQ, United Kingdom }
\affiliation {ESPCI, CNRS, F-75005 Paris, France }
\affiliation {Universit\`a di Camerino, Dipartimento di Fisica, I-62032 Camerino, Italy }
\affiliation {The University of Texas at Austin, Austin, TX 78712, USA }
\affiliation {Southern University and A\&M College, Baton Rouge, LA 70813, USA }
\affiliation {IISER-Kolkata, Mohanpur, West Bengal 741252, India }
\affiliation {National Institute for Mathematical Sciences, Daejeon 305-390, Korea }
\affiliation {RRCAT, Indore MP 452013, India }
\affiliation {Tata Institute for Fundamental Research, Mumbai 400005, India }
\affiliation {Louisiana Tech University, Ruston, LA 71272, USA }
\affiliation {SUPA, University of the West of Scotland, Paisley, PA1 2BE, United Kingdom }
\affiliation {Institute of Astronomy, 65-265 Zielona G\'ora, Poland }
\affiliation {Indian Institute of Technology, Gandhinagar Ahmedabad Gujarat 382424, India }
\affiliation {Andrews University, Berrien Springs, MI 49104, USA }
\affiliation {Trinity University, San Antonio, TX 78212, USA }
\affiliation {INFN, Sezione di Padova, I-35131 Padova, Italy }
\affiliation {University of Washington, Seattle, WA 98195, USA }
\affiliation {Southeastern Louisiana University, Hammond, LA 70402, USA }
\affiliation {Abilene Christian University, Abilene, TX 79699, USA }

\maketitle

\unnumberedsection{Introduction}
A cosmic network of strings may form as a result of phase transitions
in the early Universe~\cite{Kibble:1976sj}. When a U(1) symmetry is
broken in multiple causally disconnected spacetime regions,
one-dimensional topological defects, i.e., strings, are expected to
form~\cite{VilenkinShellard:94}. For a long time, cosmic strings were
considered candidate sources for structure formation in the early
Universe~\cite{Zeldovitch:80}. Cosmic microwave background (CMB)
experiments, however, have shown that cosmic strings can only
contribute up to a few percent of the overall anisotropies
observed~\cite{Wyman:2005tu,Bevis:2007gh,Battye:2010hg,Battye:2010xz,Ade:2013xla}. More
recently it was realized that strings can also be produced within the
framework of string theory inspired cosmological models and grow to
cosmic
scales~\cite{Linde:1993cn,Copeland:1994vg,Dvali:1994ms,Sarangi:2002yt,Jeannerot:2003qv}. Cosmic
strings produced in string theory motivated models (dubbed “cosmic
superstrings”) have received much attention since they could provide
observational signatures of string
theory~\cite{Witten:1985,Copeland:2011dx}.

Observational constraints on cosmic string models are often given as bounds
on the string tension $G\mu$ ($c=1$), where $G$ is Newton's constant
and $\mu$ the mass per unit length. Such constraints have
been derived from direct searches for line discontinuities in the
CMB temperature maps~\cite{Kaiser:1984,Lo:2005xt,Jeong:2007} and from
simulations of string-sourced CMB
anisotropies~\cite{Wyman:2005tu,Bevis:2007gh,Battye:2010hg,Battye:2010xz,Urrestilla:2011gr,Dunkley:2010ge}. These analyses, based on various assumptions about the string network, set
upper limits on $G\mu$ in the range of $10^{-7}$--$10^{-6}$. The recent
results from the Planck mission~\cite{Ade:2013xla} constrain $G\mu$ to be lower
than $1.5\times 10^{-7}$ and $3.2\times 10^{-7}$ for Nambu-Goto and
Abelian-Higgs strings, respectively.

A promising way of detecting the presence of cosmic strings and
superstrings is the gravitational wave (GW) emission from
loops~\cite{Damour:2004kw,Olmez:2010bi}. When two string segments
meet, they exchange partners or intercommute with a probability
$p$. For superstrings, the reconnection probability can be less than
unity ($10^{-4}<p<1$~\cite{Jackson:2004zg}) while field theory
simulations show that topological strings will essentially always
reconnect. This is partly due to the fact that fundamental strings
interact probabilistically. Furthermore, superstring models have extra
spatial dimensions so that even though two strings may meet in three
dimensions, they miss each other in the extra dimensions. When a string
intercommutes with itself, a closed loop breaks off. The loop
oscillates, radiates gravitationally, and decays. Cosmic string loops
can form cusps, points along the string with large Lorentz boosts,
that produce powerful bursts of gravitational
radiation~\cite{Damour:2000wa}. This Letter reports on the search for
such GW transient signatures from cosmic strings in data from the LIGO
and Virgo gravitational wave detectors.

The GW emission by cusps depends on the loop size, which is often
written as a fraction of the horizon at the time of formation $l =
\alpha t$, where $t$ is the cosmic time. Early simulations such
as Ref.~\cite{Bennett:1987vf} suggested that the size of loops is set by
gravitational backreaction and $\alpha \leq \Gamma G \mu$, where
$\Gamma \sim 50$~\cite{VilenkinShellard:94}. More recent
simulations favor cosmic string networks where the size of
loops is dictated by the large scale dynamics of the network, in which
case $\alpha \apprle
1$~\cite{Ringeval:2005kr,Blanco-Pillado:2013qja}. In this case loops
are large and they are long lived because their gravitational decay
takes many Hubble times. This Letter only reports constraints for the
small loop regime, since the large loop scenario is already well
constrained by pulsar data~\cite{Jenet:2006sv,Battye:2010xz}. We
parametrize $\alpha = \varepsilon \Gamma G \mu$ with $\varepsilon <
1$ following the convention of Ref.~\cite{Damour:2004kw}.

\begin{figure}
  \center
  \epsfig{file=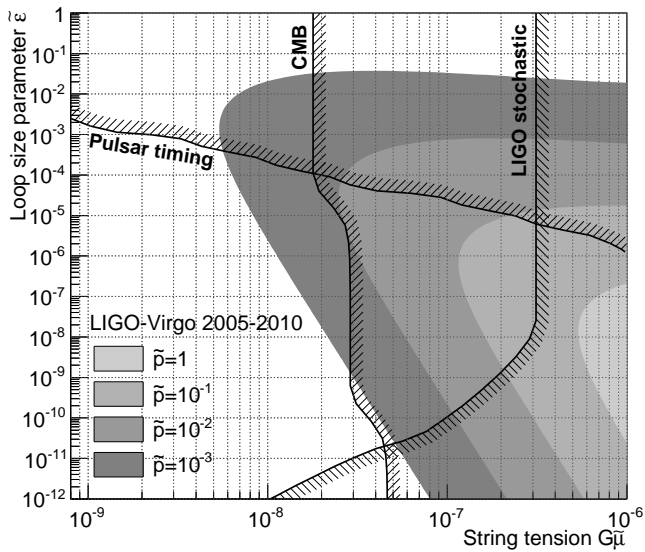,width=8.5cm,angle=0}
  \caption{Constraints on the modified cosmic string parameters
    $G\tilde{\mu}=g_1g_2^{-2/3}G\mu$,
    $\tilde{\varepsilon}=g_1^{-1}g_2^{5/3}\varepsilon$, and
    $\tilde{p}=(n_cg_1)^{-1}g_2^{-1/3}p$, where $g_1$, $g_2$ and $n_c$
    are numerical factors of $\mathcal{O}(1)$. The gray regions, in
    different shades for four reconnection probability values, are
    rejected by our analysis at a 90\% level. The black lines show
    the bounds derived from the GW stochastic background spectrum for
    $\tilde{p}=10^{-3}$ and for a small loop scenario (CMB, pulsar and
    LIGO data). The rejected region is always on the right-hand side
    of these lines.}
  \label{fig:upperlimit}
\end{figure}

Constraints on $G\mu$ were previously derived from observational limits
on the stochastic GW background expected from the incoherent
superposition of GW emission from cosmic string
loops~\cite{Olmez:2010bi}. The tightest limits were obtained, for the
case of large loops, by pulsar timing experiments~\cite{Jenet:2006sv,Battye:2010xz,Olmez:2010bi} with $G\mu
\lesssim 10^{-9}$ for $p=1$ and $G\mu \lesssim 10^{-12}$ for
$p<10^{-2}$ with $\alpha\sim0.1$. Constraints from pulsar timing
experiments were also derived for small loops~\cite{Olmez:2010bi} and
are included in Fig.~\ref{fig:upperlimit}. LIGO complemented these
results with observational limits from its own search for GW
stochastic background in the very small loop
region~\cite{Abbott:2009ws}: $G\mu \lesssim 3 \times 10^{-8}$ for
$p<10^{-3}$ and for $\varepsilon\sim10^{-11}$. Additional bounds on
the GW background can be indirectly derived from
CMB~\cite{Smith:2006nka} data and big-bang nucleosynthesis
constraints~\cite{Cyburt:2004yc}. At the epochs of last scattering and
big-bang nucleosynthesis, the energy density of the GW background must be sufficiently
small so as not to distort the CMB fluctuations or affect the
abundances of primordial elements. The CMB bound is shown in
Fig.~\ref{fig:upperlimit} and, until the present publication, offered
the best limit on $G\mu$ for intermediate values
$\varepsilon$. However this indirect limit only applies to
gravitational waves generated prior to decoupling, while LIGO and
pulsar timing data are also sensitive to later production of GWs.

\unnumberedsection{GW bursts from cosmic string cusps}
This Letter presents a different approach to constrain cosmic string
parameters, with a targeted search for transient cusp signatures in
LIGO and Virgo data. This approach was previously tested
in Ref.~\cite{Abbott:2009rr} over a short period of about 2 weeks of
live time with detectors less sensitive by about a factor of 2.
For this work we have analyzed all available LIGO and Virgo data
collected between 2005 and 2010, at design sensitivity. Moreover, the
search pipeline includes new postprocessing techniques, described in
this Letter, offering the tightest observational constraints achievable
from first generation ground-based GW interferometers.

The possibility of direct detection of GW bursts from cosmic string
cusps was first suggested in 2000 by Berezinsky \textit{et al.}~\cite{Berezinsky:2000vn}. Shortly after, Damour and Vilenkin
showed that the stochastic GW background generated by oscillating
loops is strongly non-Gaussian~\cite{Damour:2000wa}. Occasional sharp
bursts of GWs produced by cusps are expected to stand out above the
stochastic
background~\cite{Damour:2000wa,Damour:2001bk,Damour:2004kw}. Damour
and Vilenkin predict that the GW burst signal produced by cusps is
linearly polarized and the expected waveform in the frequency domain
is $h_{cusp}(f)=Af^{-4/3}$ with an exponential decay that sets on at
frequency $f_h$. The signal amplitude $A$ is determined by the string
tension, the loop size, and the propagation distance. The high
frequency cutoff $f_h$ is determined by the size of the loop and the
angle between the line of sight and the direction of the moving
cusp. It can be arbitrarily large; therefore, we take $f_h$ to be a
free parameter. Here we report on a direct search for these signatures
in LIGO and Virgo, and constrain a yet unexplored region of the string
parameter space $(G\mu,\varepsilon,p)$.

\unnumberedsection{The search}
The LIGO-Virgo detector network~\cite{Accadia:2012zzb,Abbott:2007kv}
is composed of four kilometer-scale Michelson interferometers: H1
(4~km) and H2 (2~km) share the same location at Hanford, Washington,
USA, L1 (4~km) is in Livingston Parish, Louisiana, USA, and
V1 (3~km) is located near Pisa, Italy. We analyze data collected between
November 2005 and October 2010, at times when at least two detectors
were operating simultaneously in stable conditions. This corresponds
to a total of 625 days of observation time.

The search for GW bursts from cosmic strings begins with a
matched-filter analysis of strain data from each detector
separately~\cite{Siemens:2006vk}. It consists of projecting the
whitened data onto an overpopulated~\footnote{the maximal mismatch
  between two consecutive templates is 0.001.} template
bank defined by a set of cusp waveforms with a high-frequency cutoff
spanning from 75 up to 8192~Hz. This procedure results in a
time series for the signal-to-noise ratio (SNR) for each template. An
event is identified when the SNR~$>$~3.6 and only the template with the
largest SNR is retained when several templates are triggered at the
same time. A set of five variables is used to characterize an
event. The event time $t_e$ and the SNR $\rho$ are determined by the
point where the SNR time series is maximum. The triggered template
provides the high-frequency cutoff $f_h$ and the amplitude $A$. In
addition, a $\chi^2$ parameter can be computed to characterize the
match between the event and the signal waveform in the time
domain~\cite{Allen:2004gu}.

Many transient noise events can mimic the properties of a GW burst
from a cusp. They constitute the background of our search and reduce
our chances of detecting weak signals. A fraction of these events is
removed by data quality vetoes specific to each
interferometer~\cite{Christensen:2010zz,Aasi:2012wd}. A stronger
handle on background is the requirement of simultaneous detection
in two or more interferometers. The central time of the
single-detector events must lie within a time window
sufficiently large to take into account the maximum light travel time
between detectors, the signal duration, and the timing uncertainty. For
each pair of detectors, a coincident event is characterized by three
variables: $(\delta t_e, r\!A, r\!f_h)$ where $\delta$ is used for the
difference and $r$ for the ratio between the two detectors of the pair.

To discriminate true signals from background events, we apply the
multivariate technique described in Ref.~\cite{Cannon:2008zz}, which uses
a set of simulated GW events and typical noise events to statistically
infer the probability for a candidate to be signal or noise. Given a
set of parameters $\vec{x}$ describing an event $E$, a likelihood
ratio can be computed and used to rank the events:
\begin{equation}
  \Lambda(\vec{x})= \frac{P(\vec{x}|E \in S)}{P(\vec{x}|E \in B)},
\label{eq:likelihood}
\end{equation}
where $S$ and $B$ refer to the signal and background training sample,
respectively.  The background sample is obtained by artificially
time shifting the single-detector triggers prior to the coincidence
search. The signal sample is generated by injecting simulated cosmic
string signals in the detectors' data. The simulated sources are
uniformly distributed in volume and the distribution of frequency
cutoffs $f_h$ is $dN \propto
f_h^{-5/3}df_h$~\cite{Siemens:2006vk}. The simulated signals are
injected on a time-shifted data set in order not to bias the event
ranking performed on the nonshifted data.

We parametrize an event by the coincidence
variables $\delta t_e$, $r\!A$, and $r\!f_h$ given for each of
the six possible pairs of detectors. These variables allow us to favor
signals that are coherent in the network. We also include the
single-detector SNR and $\chi^2$ parameters to discriminate genuine
signals from noise. An additional parameter is the identifier for
which set of interferometers was involved in each event, one of 11
possible combinations, to account for the different sensitivity of,
for instance, a two-detector network versus a three-detector network.
An event is therefore represented by a total of 27 variables. The
large dimensionality of the parameter space presents a computational
challenge. To obtain statistically reliable results, this method would
require very large signal and background samples, well beyond the
capabilities of present-day computers. Instead, we assume the
parameters to be uncorrelated so the likelihood ratio of
Eq.~\ref{eq:likelihood} can be factorized as
\begin{equation}
  \Lambda(\vec{x}) \approx \prod_{i=0}^{27}{\Lambda(x_i)} 
  = \prod_{i=0}^{27}{\frac{P(x_i|E \in S)}{P(x_i|E \in B)}}.
\label{eq:likelihood_product}
\end{equation}
This allows us to compute the likelihood ratio one
variable at a time. Since this estimator of $\Lambda$ neglects possible
correlations between parameters, it might result in the search being
less sensitive, compared to the idealized case where the full
27-dimensional likelihood ratio is known. In fact, we do not
perform such a factorization for the SNR and $\chi^2$ parameters
because of the strong correlation between these two
variables.

\begin{figure}
  \center
  \epsfig{file=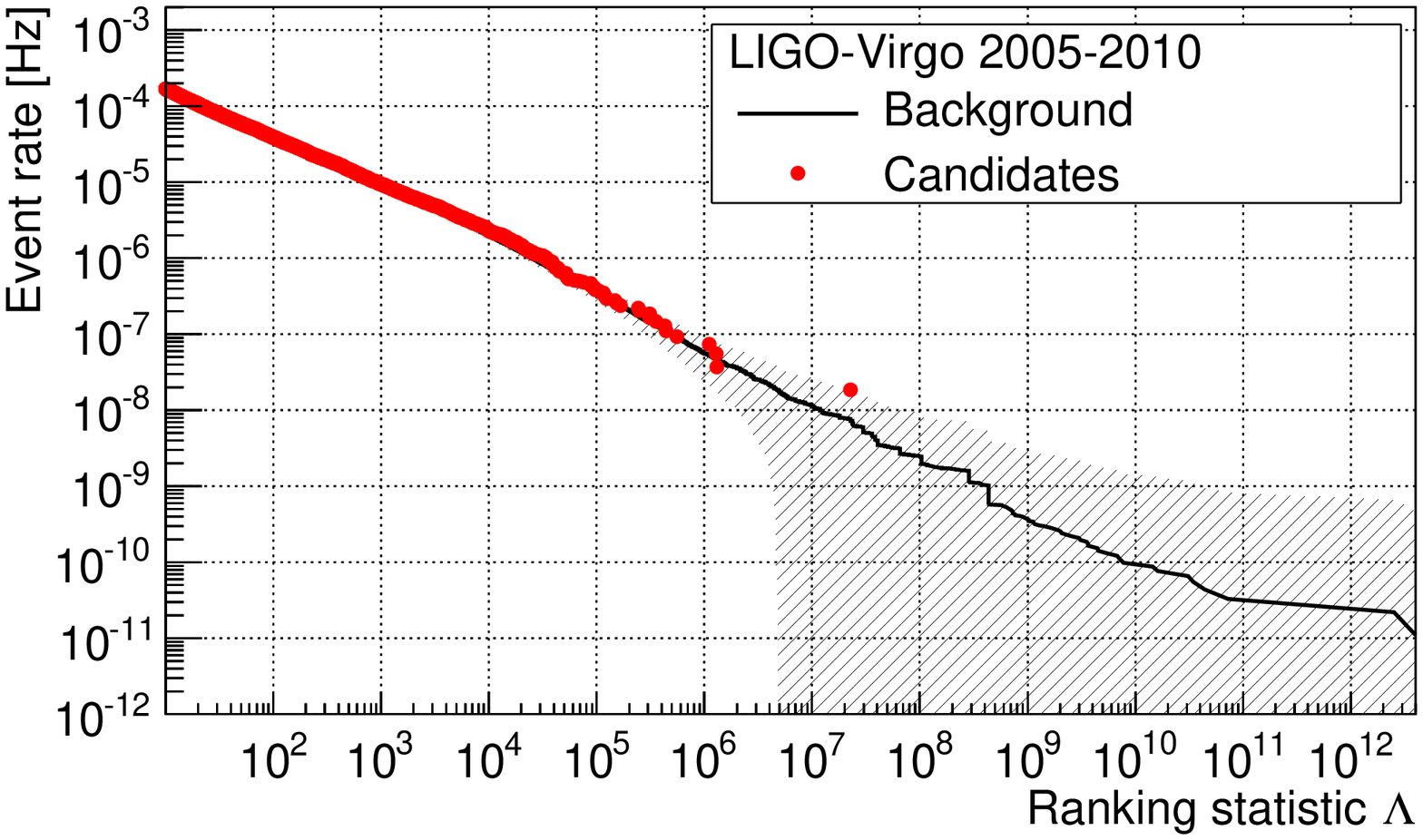,width=8.5cm,angle=0}
  \epsfig{file=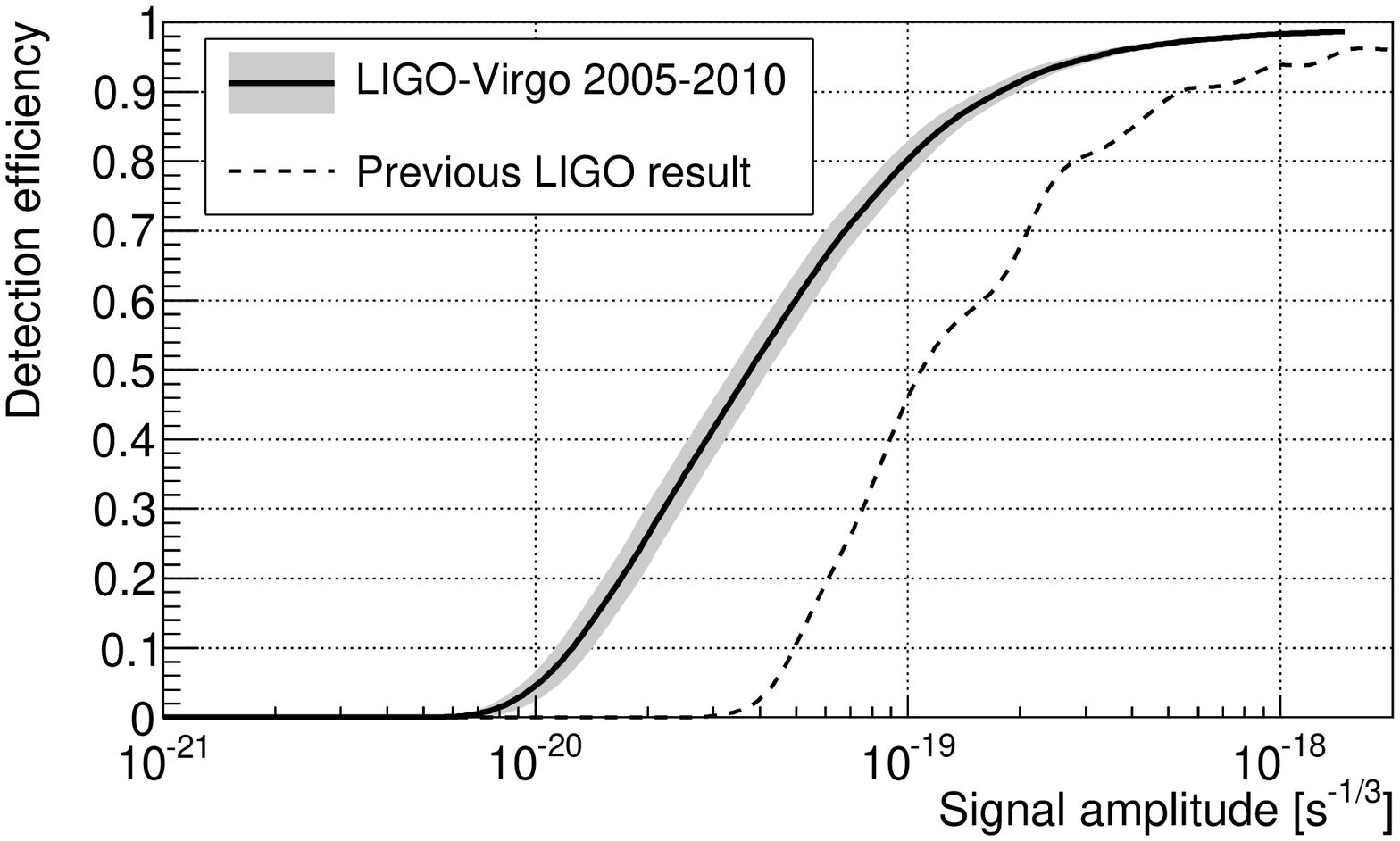,width=8.5cm,angle=0}
  \caption{In the upper plot, the red circles show the
    cumulative event rate as a function of the ranking statistic
    $\Lambda$. The black line shows the expected background of
    the search with the $1 \sigma$ statistical error represented by
    the hatched area. The highest-ranked event ($\Lambda_{h}
    \simeq 2.3\times 10^{7}$) is consistent with the background.
    The lower plot shows the sensitivity of the search as a function
    of the signal amplitude. This is measured by the fraction of
    simulated cusp events recovered with $\Lambda > \Lambda_{h}$. This
    is to be compared to the sensitivity of the previous LIGO
    search~\cite{Abbott:2009rr} represented by the dashed line.}
  \label{fig:results}
\end{figure}

\unnumberedsection{Results}
The LIGO-Virgo data set was divided into 24 time segments, which are
analyzed independently. In particular, the training sets $S$ and $B$
are generated for each segment to account for the noise
nonstationarities and the evolution of the detector sensitivities.
The principal outcome of this search is shown in
Fig.~\ref{fig:results}. The upper plot in Fig.~\ref{fig:results} shows
the combined cumulative event rate as a function of the ranking
statistic $\Lambda(\vec{x})$. The highest-ranked event of the search
occurred on May 10, 2007 at 16:27:15 UTC and is detected
simultaneously by the three LIGO interferometers. The ranking value of
this event is less than $1 \sigma$ away from the expected background
distribution from time-shifted data, shown on the same
plot. Therefore, we cannot claim this event to be a GW signal produced
by cosmic strings.

To determine the search sensitivity and derive an upper limit, about 7
million simulated cusp signals were injected into a time-shifted
data set. To avoid self-selection issues, we
use a set of injections that is independent from the $S$ sample used
to train the likelihood ranking. We run the search using the same
likelihood functions as for the nonshifted analysis and count how
many simulated signals are detected with $\Lambda$ larger
than the highest-ranked event~\cite{Brady:2004gt}. The lower plot in
Fig.~\ref{fig:results} shows the detection efficiency $e$ as
a function of the injected signal amplitude~$A$. The uncertainties
associated with the efficiency curve include binomial counting fluctuations,
calibration uncertainties, and an amplitude binning uncertainty. This
result shows that the search sensitivity has improved by a factor 3
with respect to previous LIGO results~\cite{Abbott:2009rr}. Half of
this gain is explained by the increased sensitivity of the detectors;
the rest of the gain is due to our improved statistical analysis.

A natural question we wish to answer next is what are the implications
of this nondetection for constraints in the cosmic string parameter
space. We derived model-dependent upper limits with the method
described in Ref.~\cite{Siemens:2006vk} and previously adopted
in Ref.~\cite{Abbott:2009rr}. Given the search efficiency $e(A)$, we expect
to observe an effective rate of GW bursts given by the integral over
the redshift $z$:
\begin{equation}
  \gamma(G\mu,\varepsilon,p) = \int^{\infty}_0 e(z;G\mu,\varepsilon)
  \frac{dR(z;G\mu,\varepsilon,p)}{dz}dz,
  \label{eq:loop_distribution}
\end{equation}
where $dR(z;G\mu,\varepsilon,p)$ is the cosmological rate of events
with a redshift between $z$ and $z+dz$ and is derived
in Ref.~\cite{Siemens:2006vk}. This rate relies on the generic loop density
distribution~\cite{Damour:2004kw}:
\begin{equation}
  n(l,t)=(p\Gamma G\mu)^{-1}t^{-3}\delta(l-\varepsilon \Gamma G\mu t).
  \label{eq:loop_distribution}
\end{equation}
This means that at a given cosmic time, the loop size is given by the
gravitational backreaction (the $\delta$ function) and is identical
for all loops.
Following Ref.~\cite{Damour:2001bk}, the signal amplitude is written as
\begin{equation}
  A(z;G\mu,\varepsilon)=\frac{g_1H_0^{1/3}(G\mu)^{5/3}[\varepsilon\Gamma \varphi_t(z)]^{2/3}}{(1+z)^{1/3}\varphi_r(z)},
  \label{eq:amplitude}
\end{equation}
where $g_1$ is an ignorance constant that absorbs the unknown fraction
of the loop length, which contributes to the cusp, and factors of
$\mathcal{O}(1)$. $H_0=70.1$~km~s$^{-1}$~Mpc$^{-1}$~\cite{WMAP_parameters}
is the Hubble constant. Two dimensionless cosmology-dependent
functions of the redshift $z$ enter the amplitude expression:
$\varphi_t$ and $\varphi_r$ relate the redshift and the cosmic time
$t=H_0^{-1}\varphi_t(z)$ and the proper distance $r=H_0^{-1}\varphi_r(z)$,
respectively. We use the $\varphi_r$ and $\varphi_t$ functions derived
in Appendix A of Ref.~\cite{Siemens:2006vk} for a Universe that contains
matter and radiation and includes a late-time acceleration. Those
functions are computed using the energy densities relative to the
critical density: $\Omega_m=0.279$, $\Omega_r=8.5\times 10^{-5}$, and
$\Omega_{\Lambda}=0.721$ for the matter, radiation, and cosmological constant,
respectively~\cite{WMAP_parameters}. 

Knowing how the GW amplitude $A$
scales with redshift [Eq.~(\ref{eq:amplitude})], the efficiency curve in
Fig.~\ref{fig:results} can be constructed as a function of the
redshift and parametrized with $G\mu$ and $\varepsilon$. As a result,
the parameter space $(G\mu,\varepsilon,p)$ can
be scanned and the effective rate $\gamma$ computed. The parameter
space is ruled out at a 90\% level when the effective rate exceeds
$2.303/T$, which is the expected rate from a Poisson process over an
observation time $T$. In addition to the ignorance constant $g_1$ in
Eq.~(\ref{eq:amplitude}), the $dR(z;G\mu,\varepsilon,p)$ expression
given in Ref.~\cite{Siemens:2006vk} includes two other ignorance constants:
$g_2$, and the average number of cusps per loop oscillation $n_c$. These three
constants are expected to be of $\mathcal{O}(1)$ provided cosmic
string loops are smooth. Instead of fixing these factors to 1 as it is
usually done, we choose to absorb these unknown factors in modified
cosmic string parameters: $G\tilde{\mu}=g_1g_2^{-2/3}G\mu$,
$\tilde{\varepsilon}=g_1^{-1}g_2^{5/3}\varepsilon$, and
$\tilde{p}=(n_cg_1)^{-1}g_2^{-1/3}p$.

Figure~\ref{fig:upperlimit} displays the region of the cosmic string
parameter space that is excluded by our analysis (gray-shaded
areas). For comparison, we also show limits, fixing $\tilde{p}$ at $10^{-3}$,
derived from constraints on the GW stochastic background
spectrum. These limits were computed adopting the same cosmic string
model and using the same parameters
$(G\tilde{\mu},\tilde{\varepsilon},\tilde{p})$. Our result improves
the indirect CMB bound~\cite{Smith:2006nka,Abbott:2009ws} by a factor 3 for
intermediate $\tilde{\varepsilon}$ values. It nicely complements
existing limits provided by pulsar timing experiments for large
$\tilde{\varepsilon}$~\cite{Jenet:2006sv,Battye:2010xz} and by the
LIGO stochastic search in the very small loop
regime~\cite{Abbott:2009ws}.

\unnumberedsection{Conclusion}
We found no evidence for GW bursts produced by cosmic (super)string
cusps in LIGO-Virgo data collected between 2005 and 2010. In the
absence of a detection, we place significant constraints on cosmic
string models, surpassing existing limits from CMB data. The next
generation of ground-based GW detectors will probe the cosmic string
parameter space further, including, for instance, superstring loops
with junctions~\cite{Binetruy:2010cc}, as the improved sensitivity of
Advanced LIGO~\cite{Aasi:2012wd} and Advanced Virgo~\cite{VirgoTDR}
will allow us to search for cosmic strings with an order of magnitude
lower tension.


\unnumberedsection{Acknowledgments}
The authors gratefully acknowledge the support of the United States
National Science Foundation for the construction and operation of the
LIGO Laboratory, the Science and Technology Facilities Council of the
United Kingdom, the Max-Planck-Society, and the State of
Niedersachsen/Germany for support of the construction and operation of
the GEO600 detector, and the Italian Istituto Nazionale di Fisica
Nucleare and the French Centre National de la Recherche Scientifique
for the construction and operation of the Virgo detector. The authors
also gratefully acknowledge the support of the research by these
agencies and by the Australian Research Council, 
the International Science Linkages program of the Commonwealth of Australia,
the Council of Scientific and Industrial Research of India, 
the Istituto Nazionale di Fisica Nucleare of Italy, 
the Spanish Ministerio de Econom\'ia y Competitividad,
the Conselleria d'Economia Hisenda i Innovaci\'o of the
Govern de les Illes Balears, the Foundation for Fundamental Research
on Matter supported by the Netherlands Organisation for Scientific Research, 
the Polish Ministry of Science and Higher Education, the FOCUS
Programme of Foundation for Polish Science,
the Royal Society, the Scottish Funding Council, the
Scottish Universities Physics Alliance, The National Aeronautics and
Space Administration, 
OTKA of Hungary,
the Lyon Institute of Origins (LIO),
the National Research Foundation of Korea,
Industry Canada and the Province of Ontario through the Ministry of Economic Development and Innovation, 
the National Science and Engineering Research Council Canada,
the Carnegie Trust, the Leverhulme Trust, the
David and Lucile Packard Foundation, the Research Corporation, and
the Alfred P. Sloan Foundation.

\bibliographystyle{apsrev}
\bibliography{references}

\begin{thebibliography}{46}
\expandafter\ifx\csname natexlab\endcsname\relax\def\natexlab#1{#1}\fi
\expandafter\ifx\csname bibnamefont\endcsname\relax
  \def\bibnamefont#1{#1}\fi
\expandafter\ifx\csname bibfnamefont\endcsname\relax
  \def\bibfnamefont#1{#1}\fi
\expandafter\ifx\csname citenamefont\endcsname\relax
  \def\citenamefont#1{#1}\fi
\expandafter\ifx\csname url\endcsname\relax
  \def\url#1{\texttt{#1}}\fi
\expandafter\ifx\csname urlprefix\endcsname\relax\def\urlprefix{URL }\fi
\providecommand{\bibinfo}[2]{#2}
\providecommand{\eprint}[2][]{\url{#2}}

\bibitem[{\citenamefont{Kibble}(1976)}]{Kibble:1976sj}
\bibinfo{author}{\bibfnamefont{T.~W.~B.} \bibnamefont{Kibble}},
  \bibinfo{journal}{J.Phys.A} \textbf{\bibinfo{volume}{A 9}},
  \bibinfo{pages}{1387} (\bibinfo{year}{1976}).

\bibitem[{\citenamefont{Vilenkin and Shellard}(1994)}]{VilenkinShellard:94}
\bibinfo{author}{\bibfnamefont{A.}~\bibnamefont{Vilenkin}} \bibnamefont{and}
  \bibinfo{author}{\bibfnamefont{E.}~\bibnamefont{Shellard}},
  \emph{\bibinfo{title}{Cosmic Strings and Other Topological Defects}}
  (\bibinfo{publisher}{Cambridge University Press, Cambridge, England},
  \bibinfo{year}{1994}).

\bibitem[{\citenamefont{Zeldovich}(1980)}]{Zeldovitch:80}
\bibinfo{author}{\bibfnamefont{I.~B.} \bibnamefont{Zeldovich}},
  \bibinfo{journal}{Mon. Not. R. Astron. Soc.} \textbf{\bibinfo{volume}{192}},
  \bibinfo{pages}{663} (\bibinfo{year}{1980}).

\bibitem[{\citenamefont{Wyman et~al.}(2005)\citenamefont{Wyman, Pogosian, and
  Wasserman}}]{Wyman:2005tu}
\bibinfo{author}{\bibfnamefont{M.}~\bibnamefont{Wyman}},
  \bibinfo{author}{\bibfnamefont{L.}~\bibnamefont{Pogosian}}, \bibnamefont{and}
  \bibinfo{author}{\bibfnamefont{I.}~\bibnamefont{Wasserman}},
  \bibinfo{journal}{Phys.Rev.} \textbf{\bibinfo{volume}{D 72}},
  \bibinfo{pages}{023513} (\bibinfo{year}{2005}), \eprint{astro-ph/0503364}.

\bibitem[{\citenamefont{Bevis et~al.}(2008)\citenamefont{Bevis, Hindmarsh,
  Kunz, and Urrestilla}}]{Bevis:2007gh}
\bibinfo{author}{\bibfnamefont{N.}~\bibnamefont{Bevis}},
  \bibinfo{author}{\bibfnamefont{M.}~\bibnamefont{Hindmarsh}},
  \bibinfo{author}{\bibfnamefont{M.}~\bibnamefont{Kunz}}, \bibnamefont{and}
  \bibinfo{author}{\bibfnamefont{J.}~\bibnamefont{Urrestilla}},
  \bibinfo{journal}{Phys.Rev.Lett.} \textbf{\bibinfo{volume}{100}},
  \bibinfo{pages}{021301} (\bibinfo{year}{2008}), \eprint{astro-ph/0702223}.

\bibitem[{\citenamefont{Battye et~al.}(2010)\citenamefont{Battye, Garbrecht,
  and Moss}}]{Battye:2010hg}
\bibinfo{author}{\bibfnamefont{R.}~\bibnamefont{Battye}},
  \bibinfo{author}{\bibfnamefont{B.}~\bibnamefont{Garbrecht}},
  \bibnamefont{and} \bibinfo{author}{\bibfnamefont{A.}~\bibnamefont{Moss}},
  \bibinfo{journal}{Phys.Rev.} \textbf{\bibinfo{volume}{D 81}},
  \bibinfo{pages}{123512} (\bibinfo{year}{2010}), \eprint{1001.0769}.

\bibitem[{\citenamefont{Battye and Moss}(2010)}]{Battye:2010xz}
\bibinfo{author}{\bibfnamefont{R.}~\bibnamefont{Battye}} \bibnamefont{and}
  \bibinfo{author}{\bibfnamefont{A.}~\bibnamefont{Moss}},
  \bibinfo{journal}{Phys.Rev.} \textbf{\bibinfo{volume}{D 82}},
  \bibinfo{pages}{023521} (\bibinfo{year}{2010}), \eprint{1005.0479}.

\bibitem[{\citenamefont{Ade et~al.}(2013)}]{Ade:2013xla}
\bibinfo{author}{\bibfnamefont{P.}~\bibnamefont{Ade}} \bibnamefont{et~al.}
  (\bibinfo{collaboration}{Planck Collaboration}) (\bibinfo{year}{2013}),
  \eprint{1303.5085}.

\bibitem[{\citenamefont{Linde}(1994)}]{Linde:1993cn}
\bibinfo{author}{\bibfnamefont{A.~D.} \bibnamefont{Linde}},
  \bibinfo{journal}{Phys.Rev.} \textbf{\bibinfo{volume}{D 49}},
  \bibinfo{pages}{748} (\bibinfo{year}{1994}), \eprint{astro-ph/9307002}.

\bibitem[{\citenamefont{Copeland et~al.}(1994)\citenamefont{Copeland, Liddle,
  Lyth, Stewart, and Wands}}]{Copeland:1994vg}
\bibinfo{author}{\bibfnamefont{E.~J.} \bibnamefont{Copeland}},
  \bibinfo{author}{\bibfnamefont{A.~R.} \bibnamefont{Liddle}},
  \bibinfo{author}{\bibfnamefont{D.~H.} \bibnamefont{Lyth}},
  \bibinfo{author}{\bibfnamefont{E.~D.} \bibnamefont{Stewart}},
  \bibnamefont{and} \bibinfo{author}{\bibfnamefont{D.}~\bibnamefont{Wands}},
  \bibinfo{journal}{Phys.Rev.} \textbf{\bibinfo{volume}{D 49}},
  \bibinfo{pages}{6410} (\bibinfo{year}{1994}), \eprint{astro-ph/9401011}.

\bibitem[{\citenamefont{Dvali et~al.}(1994)\citenamefont{Dvali, Shafi, and
  Schaefer}}]{Dvali:1994ms}
\bibinfo{author}{\bibfnamefont{G.}~\bibnamefont{Dvali}},
  \bibinfo{author}{\bibfnamefont{Q.}~\bibnamefont{Shafi}}, \bibnamefont{and}
  \bibinfo{author}{\bibfnamefont{R.~K.} \bibnamefont{Schaefer}},
  \bibinfo{journal}{Phys.Rev.Lett.} \textbf{\bibinfo{volume}{73}},
  \bibinfo{pages}{1886} (\bibinfo{year}{1994}), \eprint{hep-ph/9406319}.

\bibitem[{\citenamefont{Sarangi and Tye}(2002)}]{Sarangi:2002yt}
\bibinfo{author}{\bibfnamefont{S.}~\bibnamefont{Sarangi}} \bibnamefont{and}
  \bibinfo{author}{\bibfnamefont{S.~H.} \bibnamefont{Tye}},
  \bibinfo{journal}{Phys.Lett.} \textbf{\bibinfo{volume}{B 536}},
  \bibinfo{pages}{185} (\bibinfo{year}{2002}), \eprint{hep-th/0204074}.

\bibitem[{\citenamefont{Jeannerot et~al.}(2003)\citenamefont{Jeannerot, Rocher,
  and Sakellariadou}}]{Jeannerot:2003qv}
\bibinfo{author}{\bibfnamefont{R.}~\bibnamefont{Jeannerot}},
  \bibinfo{author}{\bibfnamefont{J.}~\bibnamefont{Rocher}}, \bibnamefont{and}
  \bibinfo{author}{\bibfnamefont{M.}~\bibnamefont{Sakellariadou}},
  \bibinfo{journal}{Phys.Rev.} \textbf{\bibinfo{volume}{D 68}},
  \bibinfo{pages}{103514} (\bibinfo{year}{2003}), \eprint{hep-ph/0308134}.

\bibitem[{\citenamefont{Witten}(1985)}]{Witten:1985}
\bibinfo{author}{\bibfnamefont{E.}~\bibnamefont{Witten}},
  \bibinfo{journal}{Physics Letters B} \textbf{\bibinfo{volume}{153}},
  \bibinfo{pages}{243 } (\bibinfo{year}{1985}).

\bibitem[{\citenamefont{Copeland et~al.}(2011)\citenamefont{Copeland, Pogosian,
  and Vachaspati}}]{Copeland:2011dx}
\bibinfo{author}{\bibfnamefont{E.~J.} \bibnamefont{Copeland}},
  \bibinfo{author}{\bibfnamefont{L.}~\bibnamefont{Pogosian}}, \bibnamefont{and}
  \bibinfo{author}{\bibfnamefont{T.}~\bibnamefont{Vachaspati}},
  \bibinfo{journal}{Class.Quant.Grav.} \textbf{\bibinfo{volume}{28}},
  \bibinfo{pages}{204009} (\bibinfo{year}{2011}), \eprint{1105.0207}.

\bibitem[{\citenamefont{Kaiser and Stebbins}(1984)}]{Kaiser:1984}
\bibinfo{author}{\bibfnamefont{N.}~\bibnamefont{Kaiser}} \bibnamefont{and}
  \bibinfo{author}{\bibfnamefont{A.}~\bibnamefont{Stebbins}},
  \bibinfo{journal}{Nature} \textbf{\bibinfo{volume}{310}},
  \bibinfo{pages}{391} (\bibinfo{year}{1984}).

\bibitem[{\citenamefont{Lo and Wright}(2005)}]{Lo:2005xt}
\bibinfo{author}{\bibfnamefont{A.~S.} \bibnamefont{Lo}} \bibnamefont{and}
  \bibinfo{author}{\bibfnamefont{E.~L.} \bibnamefont{Wright}}
  (\bibinfo{year}{2005}), \eprint{astro-ph/0503120}.

\bibitem[{\citenamefont{Jeong and Smoot}(2007)}]{Jeong:2007}
\bibinfo{author}{\bibfnamefont{E.}~\bibnamefont{Jeong}} \bibnamefont{and}
  \bibinfo{author}{\bibfnamefont{G.~F.} \bibnamefont{Smoot}},
  \bibinfo{journal}{The Astrophysical Journal Letters}
  \textbf{\bibinfo{volume}{661}}, \bibinfo{pages}{L1} (\bibinfo{year}{2007}),
  \eprint{astro-ph/0612706}.

\bibitem[{\citenamefont{Urrestilla et~al.}(2011)\citenamefont{Urrestilla,
  Bevis, Hindmarsh, and Kunz}}]{Urrestilla:2011gr}
\bibinfo{author}{\bibfnamefont{J.}~\bibnamefont{Urrestilla}},
  \bibinfo{author}{\bibfnamefont{N.}~\bibnamefont{Bevis}},
  \bibinfo{author}{\bibfnamefont{M.}~\bibnamefont{Hindmarsh}},
  \bibnamefont{and} \bibinfo{author}{\bibfnamefont{M.}~\bibnamefont{Kunz}},
  \bibinfo{journal}{JCAP} \textbf{\bibinfo{volume}{1112}}, \bibinfo{pages}{021}
  (\bibinfo{year}{2011}), \eprint{1108.2730}.

\bibitem[{\citenamefont{Dunkley et~al.}(2011)\citenamefont{Dunkley, Hlozek,
  Sievers, Acquaviva, Ade et~al.}}]{Dunkley:2010ge}
\bibinfo{author}{\bibfnamefont{J.}~\bibnamefont{Dunkley}},
  \bibinfo{author}{\bibfnamefont{R.}~\bibnamefont{Hlozek}},
  \bibinfo{author}{\bibfnamefont{J.}~\bibnamefont{Sievers}},
  \bibinfo{author}{\bibfnamefont{V.}~\bibnamefont{Acquaviva}},
  \bibinfo{author}{\bibfnamefont{P.}~\bibnamefont{Ade}}, \bibnamefont{et~al.},
  \bibinfo{journal}{Astrophys.J.} \textbf{\bibinfo{volume}{739}},
  \bibinfo{pages}{52} (\bibinfo{year}{2011}), \eprint{1009.0866}.

\bibitem[{\citenamefont{Damour and Vilenkin}(2005)}]{Damour:2004kw}
\bibinfo{author}{\bibfnamefont{T.}~\bibnamefont{Damour}} \bibnamefont{and}
  \bibinfo{author}{\bibfnamefont{A.}~\bibnamefont{Vilenkin}},
  \bibinfo{journal}{Phys.Rev.} \textbf{\bibinfo{volume}{D 71}},
  \bibinfo{pages}{063510} (\bibinfo{year}{2005}), \eprint{hep-th/0410222}.

\bibitem[{\citenamefont{Olmez et~al.}(2010)\citenamefont{Olmez, Mandic, and
  Siemens}}]{Olmez:2010bi}
\bibinfo{author}{\bibfnamefont{S.}~\bibnamefont{Olmez}},
  \bibinfo{author}{\bibfnamefont{V.}~\bibnamefont{Mandic}}, \bibnamefont{and}
  \bibinfo{author}{\bibfnamefont{X.}~\bibnamefont{Siemens}},
  \bibinfo{journal}{Phys.Rev.} \textbf{\bibinfo{volume}{D 81}},
  \bibinfo{pages}{104028} (\bibinfo{year}{2010}), \eprint{1004.0890}.

\bibitem[{\citenamefont{Jackson et~al.}(2005)\citenamefont{Jackson, Jones, and
  Polchinski}}]{Jackson:2004zg}
\bibinfo{author}{\bibfnamefont{M.~G.} \bibnamefont{Jackson}},
  \bibinfo{author}{\bibfnamefont{N.~T.} \bibnamefont{Jones}}, \bibnamefont{and}
  \bibinfo{author}{\bibfnamefont{J.}~\bibnamefont{Polchinski}},
  \bibinfo{journal}{JHEP} \textbf{\bibinfo{volume}{0510}}, \bibinfo{pages}{013}
  (\bibinfo{year}{2005}), \eprint{hep-th/0405229}.

\bibitem[{\citenamefont{Damour and Vilenkin}(2000)}]{Damour:2000wa}
\bibinfo{author}{\bibfnamefont{T.}~\bibnamefont{Damour}} \bibnamefont{and}
  \bibinfo{author}{\bibfnamefont{A.}~\bibnamefont{Vilenkin}},
  \bibinfo{journal}{Phys.Rev.Lett.} \textbf{\bibinfo{volume}{85}},
  \bibinfo{pages}{3761} (\bibinfo{year}{2000}), \eprint{gr-qc/0004075}.

\bibitem[{\citenamefont{Bennett and Bouchet}(1988)}]{Bennett:1987vf}
\bibinfo{author}{\bibfnamefont{D.~P.} \bibnamefont{Bennett}} \bibnamefont{and}
  \bibinfo{author}{\bibfnamefont{F.~R.} \bibnamefont{Bouchet}},
  \bibinfo{journal}{Phys.Rev.Lett.} \textbf{\bibinfo{volume}{60}},
  \bibinfo{pages}{257} (\bibinfo{year}{1988}).

\bibitem[{\citenamefont{Ringeval et~al.}(2007)\citenamefont{Ringeval,
  Sakellariadou, and Bouchet}}]{Ringeval:2005kr}
\bibinfo{author}{\bibfnamefont{C.}~\bibnamefont{Ringeval}},
  \bibinfo{author}{\bibfnamefont{M.}~\bibnamefont{Sakellariadou}},
  \bibnamefont{and} \bibinfo{author}{\bibfnamefont{F.}~\bibnamefont{Bouchet}},
  \bibinfo{journal}{JCAP} \textbf{\bibinfo{volume}{0702}}, \bibinfo{pages}{023}
  (\bibinfo{year}{2007}), \eprint{astro-ph/0511646}.

\bibitem[{\citenamefont{Blanco-Pillado
  et~al.}(2013)\citenamefont{Blanco-Pillado, Olum, and
  Shlaer}}]{Blanco-Pillado:2013qja}
\bibinfo{author}{\bibfnamefont{J.~J.} \bibnamefont{Blanco-Pillado}},
  \bibinfo{author}{\bibfnamefont{K.~D.} \bibnamefont{Olum}}, \bibnamefont{and}
  \bibinfo{author}{\bibfnamefont{B.}~\bibnamefont{Shlaer}}
  (\bibinfo{year}{2013}), \eprint{1309.6637}.

\bibitem[{\citenamefont{Jenet et~al.}(2006)\citenamefont{Jenet, Hobbs, van
  Straten, Manchester, Bailes et~al.}}]{Jenet:2006sv}
\bibinfo{author}{\bibfnamefont{F.~A.} \bibnamefont{Jenet}},
  \bibinfo{author}{\bibfnamefont{G.}~\bibnamefont{Hobbs}},
  \bibinfo{author}{\bibfnamefont{W.}~\bibnamefont{van Straten}},
  \bibinfo{author}{\bibfnamefont{R.}~\bibnamefont{Manchester}},
  \bibinfo{author}{\bibfnamefont{M.}~\bibnamefont{Bailes}},
  \bibnamefont{et~al.}, \bibinfo{journal}{Astrophys.J.}
  \textbf{\bibinfo{volume}{653}}, \bibinfo{pages}{1571} (\bibinfo{year}{2006}),
  \eprint{astro-ph/0609013}.

\bibitem[{\citenamefont{Abbott et~al.}(2009{\natexlab{a}})}]{Abbott:2009ws}
\bibinfo{author}{\bibfnamefont{B.}~\bibnamefont{Abbott}} \bibnamefont{et~al.}
  (\bibinfo{collaboration}{LIGO Scientific Collaboration, Virgo
  Collaboration}), \bibinfo{journal}{Nature} \textbf{\bibinfo{volume}{460}},
  \bibinfo{pages}{990} (\bibinfo{year}{2009}{\natexlab{a}}),
  \eprint{0910.5772}.

\bibitem[{\citenamefont{Smith et~al.}(2006)\citenamefont{Smith, Pierpaoli, and
  Kamionkowski}}]{Smith:2006nka}
\bibinfo{author}{\bibfnamefont{T.~L.} \bibnamefont{Smith}},
  \bibinfo{author}{\bibfnamefont{E.}~\bibnamefont{Pierpaoli}},
  \bibnamefont{and}
  \bibinfo{author}{\bibfnamefont{M.}~\bibnamefont{Kamionkowski}},
  \bibinfo{journal}{Phys.Rev.Lett.} \textbf{\bibinfo{volume}{97}},
  \bibinfo{pages}{021301} (\bibinfo{year}{2006}), \eprint{astro-ph/0603144}.

\bibitem[{\citenamefont{Cyburt et~al.}(2005)\citenamefont{Cyburt, Fields,
  Olive, and Skillman}}]{Cyburt:2004yc}
\bibinfo{author}{\bibfnamefont{R.~H.} \bibnamefont{Cyburt}},
  \bibinfo{author}{\bibfnamefont{B.~D.} \bibnamefont{Fields}},
  \bibinfo{author}{\bibfnamefont{K.~A.} \bibnamefont{Olive}}, \bibnamefont{and}
  \bibinfo{author}{\bibfnamefont{E.}~\bibnamefont{Skillman}},
  \bibinfo{journal}{Astropart.Phys.} \textbf{\bibinfo{volume}{23}},
  \bibinfo{pages}{313} (\bibinfo{year}{2005}), \eprint{astro-ph/0408033}.

\bibitem[{\citenamefont{Abbott et~al.}(2009{\natexlab{b}})}]{Abbott:2009rr}
\bibinfo{author}{\bibfnamefont{B.}~\bibnamefont{Abbott}} \bibnamefont{et~al.}
  (\bibinfo{collaboration}{LIGO Scientific Collaboration}),
  \bibinfo{journal}{Phys.Rev.} \textbf{\bibinfo{volume}{D 80}},
  \bibinfo{pages}{062002} (\bibinfo{year}{2009}{\natexlab{b}}),
  \eprint{0904.4718}.

\bibitem[{\citenamefont{Berezinsky et~al.}(2000)\citenamefont{Berezinsky,
  Hnatyk, and Vilenkin}}]{Berezinsky:2000vn}
\bibinfo{author}{\bibfnamefont{V.}~\bibnamefont{Berezinsky}},
  \bibinfo{author}{\bibfnamefont{B.}~\bibnamefont{Hnatyk}}, \bibnamefont{and}
  \bibinfo{author}{\bibfnamefont{A.}~\bibnamefont{Vilenkin}}
  (\bibinfo{year}{2000}), \eprint{astro-ph/0001213}.

\bibitem[{\citenamefont{Damour and Vilenkin}(2001)}]{Damour:2001bk}
\bibinfo{author}{\bibfnamefont{T.}~\bibnamefont{Damour}} \bibnamefont{and}
  \bibinfo{author}{\bibfnamefont{A.}~\bibnamefont{Vilenkin}},
  \bibinfo{journal}{Phys.Rev.} \textbf{\bibinfo{volume}{D 64}},
  \bibinfo{pages}{064008} (\bibinfo{year}{2001}), \eprint{gr-qc/0104026}.

\bibitem[{\citenamefont{Accadia et~al.}(2012{\natexlab{a}})}]{Accadia:2012zzb}
\bibinfo{author}{\bibfnamefont{T.}~\bibnamefont{Accadia}} \bibnamefont{et~al.}
  (\bibinfo{collaboration}{Virgo Collaboration}), \bibinfo{journal}{JINST}
  \textbf{\bibinfo{volume}{7}}, \bibinfo{pages}{P03012}
  (\bibinfo{year}{2012}{\natexlab{a}}).

\bibitem[{\citenamefont{Abbott et~al.}(2009{\natexlab{c}})}]{Abbott:2007kv}
\bibinfo{author}{\bibfnamefont{B.}~\bibnamefont{Abbott}} \bibnamefont{et~al.}
  (\bibinfo{collaboration}{LIGO Scientific Collaboration}),
  \bibinfo{journal}{Rept.Prog.Phys.} \textbf{\bibinfo{volume}{72}},
  \bibinfo{pages}{076901} (\bibinfo{year}{2009}{\natexlab{c}}),
  \eprint{0711.3041}.

\bibitem[{\citenamefont{Siemens et~al.}(2006)\citenamefont{Siemens, Creighton,
  Maor, Ray~Majumder, Cannon et~al.}}]{Siemens:2006vk}
\bibinfo{author}{\bibfnamefont{X.}~\bibnamefont{Siemens}},
  \bibinfo{author}{\bibfnamefont{J.}~\bibnamefont{Creighton}},
  \bibinfo{author}{\bibfnamefont{I.}~\bibnamefont{Maor}},
  \bibinfo{author}{\bibfnamefont{S.}~\bibnamefont{Ray~Majumder}},
  \bibinfo{author}{\bibfnamefont{K.}~\bibnamefont{Cannon}},
  \bibnamefont{et~al.}, \bibinfo{journal}{Phys.Rev.} \textbf{\bibinfo{volume}{D
  73}}, \bibinfo{pages}{105001} (\bibinfo{year}{2006}), \eprint{gr-qc/0603115}.

\bibitem[{Note1()}]{Note1}
Note1, \bibinfo{note}{the maximal mismatch between two consecutive templates is
  0.001.}

\bibitem[{\citenamefont{Allen}(2005)}]{Allen:2004gu}
\bibinfo{author}{\bibfnamefont{B.}~\bibnamefont{Allen}},
  \bibinfo{journal}{Phys.Rev.} \textbf{\bibinfo{volume}{D 71}},
  \bibinfo{pages}{062001} (\bibinfo{year}{2005}), \eprint{gr-qc/0405045}.

\bibitem[{\citenamefont{Christensen}(2010)}]{Christensen:2010zz}
\bibinfo{author}{\bibfnamefont{N.}~\bibnamefont{Christensen}}
  (\bibinfo{collaboration}{LIGO and Virgo Scientific Collaborations}),
  \bibinfo{journal}{Class.Quant.Grav.} \textbf{\bibinfo{volume}{27}},
  \bibinfo{pages}{194010} (\bibinfo{year}{2010}).

\bibitem[{\citenamefont{Aasi et~al.}(2012)}]{Aasi:2012wd}
\bibinfo{author}{\bibfnamefont{J.}~\bibnamefont{Aasi}} \bibnamefont{et~al.}
  (\bibinfo{collaboration}{Virgo Collaboration}),
  \bibinfo{journal}{Class.Quant.Grav.} \textbf{\bibinfo{volume}{29}},
  \bibinfo{pages}{155002} (\bibinfo{year}{2012}), \eprint{1203.5613}.

\bibitem[{\citenamefont{Cannon}(2008)}]{Cannon:2008zz}
\bibinfo{author}{\bibfnamefont{K.~C.} \bibnamefont{Cannon}},
  \bibinfo{journal}{Class.Quant.Grav.} \textbf{\bibinfo{volume}{25}},
  \bibinfo{pages}{105024} (\bibinfo{year}{2008}).

\bibitem[{\citenamefont{Brady et~al.}(2004)\citenamefont{Brady, Creighton, and
  Wiseman}}]{Brady:2004gt}
\bibinfo{author}{\bibfnamefont{P.~R.} \bibnamefont{Brady}},
  \bibinfo{author}{\bibfnamefont{J.~D.} \bibnamefont{Creighton}},
  \bibnamefont{and} \bibinfo{author}{\bibfnamefont{A.~G.}
  \bibnamefont{Wiseman}}, \bibinfo{journal}{Class.Quant.Grav.}
  \textbf{\bibinfo{volume}{21}}, \bibinfo{pages}{S1775} (\bibinfo{year}{2004}),
  \eprint{gr-qc/0405044}.

\bibitem[{WMA()}]{WMAP_parameters}
\urlprefix\url{http://lambda.gsfc.nasa.gov/product/map/dr3/parameters_summary.cfm}.

\bibitem[{\citenamefont{Binetruy et~al.}(2010)\citenamefont{Binetruy, Bohe,
  Hertog, and Steer}}]{Binetruy:2010cc}
\bibinfo{author}{\bibfnamefont{P.}~\bibnamefont{Binetruy}},
  \bibinfo{author}{\bibfnamefont{A.}~\bibnamefont{Bohe}},
  \bibinfo{author}{\bibfnamefont{T.}~\bibnamefont{Hertog}}, \bibnamefont{and}
  \bibinfo{author}{\bibfnamefont{D.~A.} \bibnamefont{Steer}},
  \bibinfo{journal}{Phys.Rev.} \textbf{\bibinfo{volume}{D 82}},
  \bibinfo{pages}{126007} (\bibinfo{year}{2010}), \eprint{1009.2484}.

\bibitem[{\citenamefont{Accadia et~al.}(2012{\natexlab{b}})}]{VirgoTDR}
\bibinfo{author}{\bibfnamefont{T.}~\bibnamefont{Accadia}} \bibnamefont{et~al.},
  \emph{\bibinfo{title}{{Virgo Document VIR-0128A-12}}}
  (\bibinfo{year}{2012}{\natexlab{b}}),
  \urlprefix\url{https://tds.ego-gw.it/ql/?c=8940}.

\end{thebibliography}

\end{document}